\documentclass{article}
\usepackage{amsmath,graphicx}
\usepackage[a4paper]{geometry}
\usepackage{xcolor}
\usepackage{amsfonts,authblk}
\usepackage{hyperref}

\title{Advantages and limitations of channel multiplexing for discrete-variable entanglement-based quantum key distribution}

\author[1]{Indranil Maiti\thanks{\href{mailto:indra@doktorant.umk.pl}{Corresponding author: indra@doktorant.umk.pl}}}
\author[1]{Miko\l aj Lasota}
\affil[1]{Faculty of Physics, Astronomy and Informatics, Nicolaus Copernicus University, Grudziadzka 5, 87-100 Toru\'{n}, Poland}

\begin{document}

\maketitle

\begin{abstract}
Typically practical realizations of discrete-variable quantum key distribution (QKD) protocols, based on exchanging single-photon signals between the trusted parties, can provide its users with only very low key generation rates. One of the potential solutions for this problem, that can be adapted for the case of entanglement-based QKD schemes using broadband photon-pair sources, is to utilize wavelength-division-multiplexing (WDM) modules in order to split the photons with different wavelengths to separate detection channels and generate multiple keys in parallel. Here, we theoretically investigate this idea in the case of pulsed laser used to pump spontaneous parametric down-conversion source. We optimize the effective phase-matching function width of the nonlinear crystal, and the intensity and duration of the pump laser pulses in order to maximize the advantage of the overall key generation rate provided by the WDM-based QKD scheme over the traditional no-WDM scenario. The results of our analysis show that the considered method can significantly accelerate the production of cryptographic keys, but proper optimization of the photon-pair source is needed to exploit its full potential.

\end{abstract}

\section{Introduction}
\label{Sec:Introduction}

Quantum key distribution (QKD) is a method to generate cryptographic keys between distant trusted parties \cite{Pirandola2020}. Contrary to the classical protocols, the security of which relies on uncertain assumptions on the computational power accessible to a potential eavesdropper, the security of QKD schemes is based on fundamental laws of nature. Unfortunately, numerous imperfections of realistic setup elements required for the implementation of QKD protocols impose strong limitation on their performance, both in terms of maximal security distance and the obtainable key generation rate \cite{Brassard2000}. The latter can be particularly low in the case of discrete-variable (DV) protocols, that require using single photons or their approximations in the form of weak coherent pulses as information carriers. On the other hand DV QKD schemes are capable of distributing secret keys on several hundreds of kilometers \cite{Zhou2023,Liu2023}. It is considerably longer than in the case of their continuous-variable counterparts \cite{Zhang2020}, which are more vulnerable to photon loss, particularly in the case of entanglement-based setup configuration \cite{Lasota2023}. 

Among the sources of photon pairs, required for the implementation of entanglement-based versions of DV QKD protocols, the most popular are the devices based on the spontaneous parametric down-conversion (SPDC) process \cite{Louisell1961,Burham1970}. Their advantages include high generation and collection efficiency \cite{Pomarico2012,Ramelow2013,Kaneda2016}, high quality of the emitted photons \cite{Fasel2004,Bock2016}, good performance in room temperature, high versatility in terms of properties of the produced photons and relatively low construction cost. On the other hand, this type of source is probabilistic, meaning that the number of photon pairs generated from a single pump pulse follows some statistics, which is thermal in the case of single-mode SPDC process and gradually changes toward Poissonian when the number of contributing modes grow \cite{Kok2000,Helwig2009}. Multipair emission events have negative influence on the QKD security, as has the effect of temporal broadening of SPDC photons, taking place during their propagation in dispersive media due to their relatively broad spectrum \cite{Sedziak2017,Lasota2020}. 

However, the aforementioned disadvantages of SPDC sources can be turned into an actual advantage. To achieve this, both legitimate parties have to divide the spectrum of the received photons into several ranges and direct the photons from different ranges into separate detection systems. If the spectral correlation between the photons belonging to the same SPDC pair is relatively strong it can be possible to define multiple pairs of correlated systems to which these photons could go with very high probability. 

To understand how this idea can help Alice and Bob with secure key generation, let us imagine that two pairs of photons are produced by the SPDC source from a single pump pulse, and the measurement setups utilized by the trusted parties  exhibit perfect photon number resolution. Let us first focus on the case when all of the photons entering Alice’s laboratory are directed to a single detection system, and similarly for Bob. For a given event to be accepted in the key generation process only one out of two photons, send from the source to any of the parties, can survive the travel, while the other one has to be lost. The probability for the two surviving photons to belong to the same pair is exactly the same as the probability for them to be uncorrelated. Since the detection results coming from uncorrelated photon pairs produce errors in the raw key in $50\%$ of cases, the overall error rate originating from the double-pair emission events is $25\%$. It can become much lower if the trusted parties had the ability to direct the photons from different spectral ranges into separate measurement setups. In the strong spectral correlation case if one photon from an SPDC pair enters a given  detection system on Alice's side, it is possible to predict with nearly $100\%$ accuracy to which of Bob's detection systems the corresponding photon would go. Then, the trusted parties can define many pairs of such correlated detection systems and accept only the measurement results obtained in them. While almost all the pairs of correlated SPDC photons can be accepted by such a scheme, a pair of uncorrelated photons can go to many different combinations of detection systems, including uncorrelated ones, and thus be rejected. Therefore, separating the spectrum of the strongly correlated SPDC pairs into several ranges and directing them to different detection systems, can be used to introduce an additional error rejection technique to the QKD setup. 

Moreover, there is another strong benefit from such setup modification. Coming back to the above example, let us consider the situation in which all the four photons survived their travel to the trusted parties. If the traditional QKD setup is used, such an event would be automatically discarded from the key due to multiple clicks registered in both detection systems. On the contrary, when the spectrum is divided and photons with different wavelengths are sent to separate measurement setups, it can happen that the trusted parties observe single clicks in two pairs of correlated setups and can generate two bits of the raw key from a single use of the SPDC source, instead of only one. Taking this into account and realizing that the additional error rejection technique mentioned above allows Alice and Bob to increase the power of the SPDC source, leading to the generation of many pairs of photons in a single attempt, it is not difficult to realize that the idea of dividing the spectrum of strongly correlated pairs of photons and directing them to separate detection systems can be utilized to produce many cryptographic keys in parallel. If these keys are combined together at the end, the resulting overall key generation rate can become significantly higher than the rate obtainable from the traditional QKD setup configuration.

This idea has been recently suggested by Pseiner {et al.} \cite{Pseiner2021}, who experimentally investigated the possibility to increase the achievable key generation rate in QKD system based on SPDC source pumped with continuous-wave (CW) laser. In this work they utilized wavelength-division-multiplexing (WDM) modules to direct SPDC photons into separate detection setups. Such devices are one of the standard elements of modern classical communication systems \cite{AgrawalBook}. In the context of QKD they have been studied in several papers exploring the possibility to perform simultaneous quantum and classical communication with the use of single fiber link \cite{Townsend1997a,Eraerds2010,Qi2010} and quantum-classical communication networks \cite{Townsend1997b,Sun2018,Dynes2019}.

In this work we adapt the idea introduced in Ref.\,\cite{Pseiner2021} to analyze the possibility for increasing the key generation rate produced from the entanglement-based version of the BB84 protocol \cite{Bennett1984}, realized with pulsed-pump SPDC source, when using WDM modules. According to our knowledge it is the first comprehensive theoretical study of this method. In order to properly assess its potential advantage, we numerically optimize the parameters of the SPDC source, including the effective-phase matching function (EPMF) width of the nonlinear crystal, and the duration and intensity of the pump laser pulses, for various numbers of WDM channel pairs utilized by the trusted parties for QKD. We compare the results with the maximal key generation rate that can be obtained in traditional setup configuration, without WDM modules. Although not completely neglected in the QKD literature, the problem of source optimization has not been studied to sufficient extent, especially since the existing works suggest its high importance for maximizing the performance of QKD systems \cite{Sedziak2017,Lasota2020}. Therefore, we consider it to be a valuable element of our analysis. We also investigate different properties of the WDM channels, including their width, separation and transmission profile. We analyze both the cases of SPDC source producing strongly negative and strongly positive spectral correlation between the pairs of generated photons. The presented results show that the overall key rate can indeed be significantly increased using the investigated method, but the achievable advantage strongly depends on the properties of the utilized SPDC source.  

The paper is organized as follows. In Sec.\ref{Sec:Assumptions} we specify the considered setup configuration and discuss our assumptions on its main properties. The QKD security analysis is provided in Sec.\ref{Sec:Theory}. Next, in Sec.\ref{Sec:Results}, we present the results of numerical simulations, done with the use of security formulas derived in Sec.\ref{Sec:Theory}. Finally, in Sec.\ref{Sec:Summary} we  summarize the main conclusions stemming from our investigation, discuss the advantages and limitations of the presented method to increase the key generation rate, and briefly present our future plans related to this research.

\section{Setup and assumptions}
\label{Sec:Assumptions}

\begin{figure}[tb]
\centering
\includegraphics[width=0.8\linewidth]{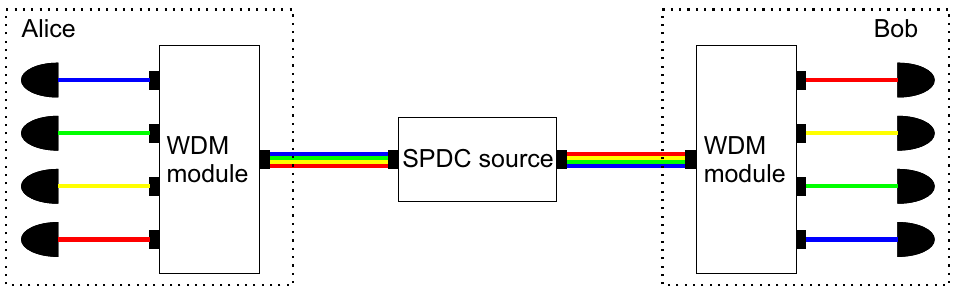}
\caption{Schematic diagram of the considered setup.}
\label{fig:setup}
\end{figure}

In this work we consider entanglement-based version of the BB84 protocol, with information encoded in photon polarization. A schematic diagram of the setup is presented in Fig.\ref{fig:setup}. We assume that the pulsed-pump type II SPDC source, located in-between Alice and Bob, produces two-photon polarization states $|\psi^-\rangle=\left(|HV\rangle-|VH\rangle\right)/\sqrt{2}$. The spectral biphoton wavefunction can be conveniently approximated by \cite{Lutz2014,Gajewski2016}:
\begin{equation}
    f(\omega_s,\omega_i)\approx\sqrt{\frac{2\tau_p}{\pi\sigma_{cr}}}\times\exp\left[-\frac{(\omega_s-\omega_i)^2}{\sigma_{cr}^2}-\frac{(\omega_s+\omega_i)^2\tau_p^2}{4}\right],
    \label{eq:wavefunction}
\end{equation}
where $\omega_s$ and $\omega_i$ denote the angular frequency detunings of signal and idler photon, respectively, from their central values, $\tau_p$ is the pump laser pulse duration, and $\sigma_{cr}$ is the EPMF width, characterizing the nonlinear crystal in which the SPDC process takes place. The parameters $\tau_p$ and $\sigma_{cr}$ are convenient from experimental point of view, since they can be changed independently from each other, by modifying the pump laser and non-linear crystal utilized by the source, respectively. However, in some situations it is beneficial to use the quantities describing the properties of the generated pairs of photons, namely their spectral bandwidth, $\sigma$, and the spectral correlation coefficient, $\rho$. The relationship between the two sets of parameters mentioned above is as follows \cite{Sedziak2019}:
\begin{equation}
    \sigma=\frac{\sqrt{4+\sigma_{cr}^2\tau_p^2}}{2\sqrt{2}\tau_p},
\end{equation}
\begin{equation}
    \rho=\frac{4-\sigma_{cr}^2\tau_p^2}{4+\sigma_{cr}^2\tau_p^2}.
\end{equation}

\begin{figure}[tb]
\centering
\includegraphics[width=\linewidth]{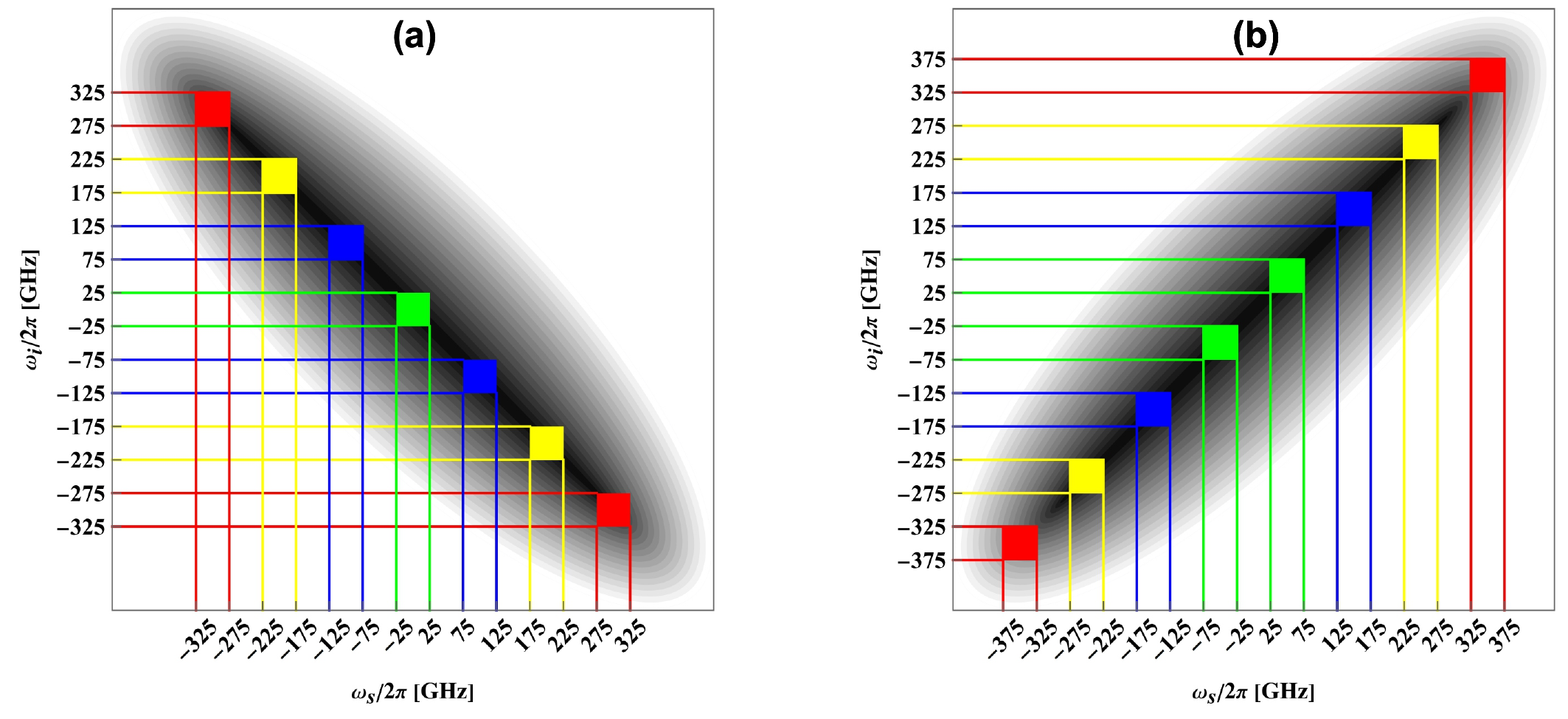}
\caption{Examples of the spectral wavefunction for pairs of photons produced by the SPDC source in the case of (a) negative, (b) positive spectral correlation generated between them. The multicoloured squares indicate a grid of (a) seven or (b) eight WDM channel pairs, with rectangular transmission profiles, that can be used for separate key generation processes.}
\label{fig:spectrum}
\end{figure}

Without losing generality of the presented consideration we assume that the signal (idler) photons are always sent to Alice (Bob). At their arrival to the trusted parties' laboratories they pass through WDM modules, which direct them to different output channels, depending on their wavelengths. Finally, their polarizations are measured at the exit of the output channels. As already mentioned in Sec.\,\ref{Sec:Introduction}, if the spectral correlation between the SPDC photons is relatively strong, $|\rho|\approx1$, detecting one of them in a particular output channel by Alice allows her to predict with high probability the output channel of Bob's setup to which the corresponding photon from the same pair would go. This is illustrated in Fig.\ref{fig:spectrum}, where we sketch the spectral wavefunction describing photon pairs generated by an SPDC source for the so-called negative spectral correlation scenario, ($\rho<0$) -- panel (a) -- and the positive spectral correlation scenario ($\rho>0$) -- panel (b). For both these cases an example of a set of correlated WDM channels that the trusted parties can use is also shown. It is assumed that only if coincident clicks are observed at the outputs of any of the chosen channel pairs, Alice and Bob accept such event for the key generation process. On the contrary, coincident clicks from any other combination of WDM channels is automatically discarded due to high probability that they are caused by uncorrelated photons.

Various WDM channel transmission profiles can be found in the literature, depending particularly on the type of optical filters utilized by the WDM module \cite{Keiser1999}. In our investigation we consider two approximations of such profile, given by the rectangular and Gaussian functions. In the case of rectangular profile the spectral wavefunction of the photon entering a given WDM channel, centered at $\omega_0$ and with $\Delta\omega$ width, is modified by the following factor:
\begin{equation}
    F_{\omega_0}^\mathrm{rect}(\omega)=\left\{\begin{array}{cl}
        1; &  \omega_0-\Delta\omega/2\leq\omega\leq\omega_0+\Delta\omega/2\\
        0; &  \mathrm{otherwise}
    \end{array}\right..
    \label{eq:rectchannel}
\end{equation}
On the other hand, for Gaussian-shaped profile, parametrized by the standard deviation $\sigma_{\!f}$, we have:
\begin{equation}
     F_{\omega_0}^\mathrm{Gauss}(\omega)=\exp\left[-\frac{(\omega-\omega_0)^2}{2\sigma_{\!f}^2}\right].
     \label{eq:Gausschannel}
\end{equation}
We also assume that the separation between the neighboring WDM channels in terms of angular frequency is fixed to $\omega_\mathrm{sep}=2\pi\times100\,\mathrm{GHz}$, which is typical value for a standard WDM channel grid. The slopes of realistic WDM channel transmission functions are typically much steeper than the Gaussian function, but not exactly vertical as in the rectangular one. Thus, the presented results of key generation rate calculation can be seen as the lower and upper bound, respectively, for the values that can be obtained when using practical WDM modules.

In order to properly describe the sets of channel pairs considered in our analysis let us introduce the channel pair identification number $n$ and the total number of WDM channel pairs utilized by the trusted parties, $N$. We assume that if the spectral correlation generated between the photon pairs produced by the SPDC source is negative, as in Fig.\ref{fig:spectrum}\,(a), then the central angular frequency detunings of the considered channel pairs are given by the set
\begin{equation}
    \left\{\omega_{s0}^{(n,-)},\omega_{i0}^{(n,-)}\right\}=\left\{-\frac{n}{2}\omega_\mathrm{sep},\frac{n}{2}\omega_\mathrm{sep}\right\}.
    \label{eq:omegasetnegative}
\end{equation}
If $N$ is odd, Alice and Bob utilize the channel pairs with even identification numbers ranging from $-(N-1)$ to $N-1$. On the contrary, if $N$ is even the channel pairs with odd identification numbers between $-N$ and $N$ are used. For example the channel pairs illustrated in Fig.\,\ref{fig:spectrum}\,(a) have the following identification numbers according to our notation: -6, -4, -2, 0, 2, 4, 6 (counting from left top red to right bottom red square). With such channel numbering system we fix the center of the WDM channel grid to coincide with the center of the biphoton wavefunction, regardless of how many channel pairs are utilized by the trusted parties. This choice is reasonable since it maximizes the obtainable overall key generation rate for any fixed value of $N$. It can be also seen that the individual key generation rates provided by the channel pairs with numbers $-n$ and $n$ (represented by the squares of the same colors in Fig.\,\ref{fig:spectrum}) are equal to each other. Similar channel numbering system is used for the case of positive spectral correlation produced between the SPDC photons, but in this scenario the set of central angular frequency detunings considered in our analysis is modified to 
\begin{equation}
    \left\{\omega_{s0}^{(n,+)},\omega_{i0}^{(n,+)}\right\}=\left\{\frac{n}{2}\omega_\mathrm{sep},\frac{n}{2}\omega_\mathrm{sep}\right\}.
    \label{eq:omegasetpositive}
\end{equation}
In the example of Fig.\,\ref{fig:spectrum}\,(b), where eight channel pairs are pictured, their central angular frequency detunings are given by Eq.\, (\ref{eq:omegasetpositive}), with $n$ taking the values from the following set: \{-7, -5, -3, -1, 1, 3, 5, 7\}.

Since the number of parameters describing the SPDC source and WDM modules, considered in this paper, is already quite high, we limit our analysis to the case of ideal photon-number-resolving (PNR) detectors utilized by Alice and Bob, and lossy, but noiseless channels connecting them with the source. Therefore, the presented results are most relevant for medium-distance QKD implementations, where the probability for the signal emitted by the source to reach the measurement systems of the trusted parties is much larger than the overall probability to register noise, while the probability for Alice and/or Bob to observe more than one photon in a single WDM channel is negligible. We use the symbols $T_A$ and $T_B$ to denote the channels' transmittance. While the theoretical analysis is performed for the general case, in numerical simulations we focus solely on the symmetric situation, in which $T_A=T_B\equiv T$. The extension of this analysis to the more practical cases of noisy QKD setup and binary detectors is going to be presented in the follow-up to this paper, which is currently in preparation.


\section{Key generation rate}
\label{Sec:Theory}

The main quantity utilized for the security evaluation of QKD protocols is the key generation rate, defined as the number of secure key bits produced by the trusted parties per one use of the source. In general it can be expressed as \cite{Scarani2009}
\begin{equation}
    K = p_\mathrm{sift}p_\mathrm{acc}\left[I(A:B)-\mathrm{max}(I_{EA},I_{EB})\right],
    \label{eq:keyratemain}
\end{equation}
where $p_\mathrm{sift}$ is the so-called sifting probability, $p_\mathrm{acc}$ is the probability for Alice and Bob to accept the outcome of their measurements for the key generation process, $I(A:B)$ is the mutual information between the trusted parties and $I_{EA}$ ($I_{EB}$) denotes the information gained by the eavesdropper on Alice's (Bob's) version of the raw key. For protocols with binary encoding, like BB84, $I(A:B)=1-h(Q)$, where $Q$ is the so-called quantum bit error rate (QBER) and $h(Q)$ denotes binary Shannon entropy. The value of $p_\mathrm{sift}$ is equal to the probability for Alice and Bob to choose matching measurement bases, which allow them to obtain correlated detection results. Otherwise the event has to be discarded from the raw key. This is done during the postprocessing stage of the protocol, in the procedure called sifting \cite{Scarani2009}. In this work we focus on the traditional version of the BB84 protocol, with symmetric basis choice, in which case $p_\mathrm{sift}=1/2$. This assumption does not change the results of our calculation in any meaningful way, since it merely introduces a constant scaling factor for all the calculated key rates.

Let us now focus on one pair of correlated WDM channels utilized by Alice and Bob, centered at $\omega_{s0}$ and $\omega_{i0}$, respectively. If the trusted parties employ ideal PNR detectors for their measurements, they can immediately uncover large-pulse attacks \cite{Lutkenhaus1999}, that could potentially be performed by Eve. Therefore, they can automatically remove all the double click\footnote{Here we would like to underline that by double click we understand only a pair of clicks registered in the two detectors belonging to the same detection system. In particular, clicks registered in two separate detection systems when multiple keys are being produced in parallel is not a double click event} events from the key generation process and accept only the events when both of them received exactly one click in one of their detectors. If we denote by $p(i_H,i_V;j_H,j_V)$ the joint probability for Alice to measure $i_H$ and $i_V$ photons in her detectors for $H$- and $V$-polarized light, and for Bob to measure $j_H$ and $j_V$ photons in his detectors for $H$- and $V$- polarized light, respectively, then in this case
\begin{equation}    p_\mathrm{acc}=p(1,0;1,0)+p(1,0;0,1)+p(0,1;1,0)+p(0,1;0,1).
\label{eq:paccpnr}
\end{equation}
Furthermore, since the pairs of photons produced by the source are supposed to be anti-correlated in polarization, the QBER can be calculated as
\begin{equation}
    Q=\frac{p(1,0;1,0)+p(0,1;0,1)}{p_\mathrm{acc}}.
\end{equation}


In the entanglement-based version of the BB84 protocol photons belonging to different SPDC pairs are uncorrelated. Therefore, Eve cannot gain any information on the polarization state of the photon detected by either Alice or Bob by performing photon-number-splitting (PNS) attacks \cite{Brassard2000}. In this case the upper bound for the information on the raw key gained by her from general attacks on the BB84 protocol reads \cite{Kraus2005,Renner2005}
\begin{equation}
    I_{EA}=I_{EB}=h(Q).
    \label{eq:iae}
\end{equation}

Calculating the key generation rate requires expressing the probabilities $p(i_H,i_V;j_H,j_V)$ in terms of the setup parameters. To this end let us first denote the probability for both photons belonging to a single SPDC pair to enter the two considered WDM channels by $p_{++}$. If the channels have rectangular transmission profile, given by Eq.\,(\ref{eq:rectchannel}), then
\begin{equation}
    p_{++}^\mathrm{rect} = \int_{\omega_{s0}-\Delta\omega/2}^{\omega_{s0}+\Delta\omega/2} \mathrm{d}\omega_s \int_{\omega_{i0}-\Delta\omega/2}^{\omega_{i0}+\Delta\omega/2} \mathrm{d}\omega_i\;|f(\omega_s,\omega_i)|^2, 
     \label{eq:pplusplusrect}
\end{equation}
where the spectral biphoton wavefunction is given by the formula (\ref{eq:wavefunction}). In the case of Gaussian transmission profile (\ref{eq:Gausschannel}) the aforementioned probability transforms into
\begin{equation}
    p_{++}^\mathrm{Gaus} = \int_{-\infty}^{+\infty} \mathrm{d}\omega_s \int_{-\infty}^{+\infty} \mathrm{d}\omega_i\;|f(\omega_s,\omega_i)|^2\times\,\mathrm{exp}\left[-\frac{(\omega_s-\omega_{s0})^2+(\omega_i-\omega_{i0})^2}{\sigma_f^2}\right].
    \label{eq:pplusplusGaus}
\end{equation}

Analogously we define the probabilities that only Alice's photon enters her WDM channel, only Bob's photon enters his WDM channel and none of the photons enter the two channels by $p_{+-}$, $p_{-+}$ and $p_{--}$, respectively. They read:
\begin{equation}
    p_{+-}^\mathrm{rect} = \int_{\omega_{s0}-\Delta\omega/2}^{\omega_{s0}+\Delta\omega/2} \mathrm{d}\omega_s\int_{-\infty}^{+\infty} \mathrm{d}\omega_i\;|f(\omega_s,\omega_i)|^2-p_{++}^\mathrm{rect} , 
    \label{eq:pplusminusrect}
\end{equation}
\begin{equation}
 p_{-+}^\mathrm{rect} = \int_{-\infty}^{+\infty} \mathrm{d}\omega_s \int_{\omega_{i0}-\Delta\omega/2}^{\omega_{i0}+\Delta\omega/2} \mathrm{d}\omega_i\;|f(\omega_s,\omega_i)|^2-p_{++}^\mathrm{rect}, 
 \label{eq:pminusplusrect}
\end{equation}
\begin{equation}
 p_{--}^\mathrm{rect} = 1-p_{+-}^\mathrm{rect}-p_{-+}^\mathrm{rect}-p_{++}^\mathrm{rect},
 \label{eq:pminusminusrect}
\end{equation}
\begin{equation}
    p_{+-}^\mathrm{Gaus} = \int_{-\infty}^{+\infty} \mathrm{d}\omega_s \int_{-\infty}^{+\infty} \mathrm{d}\omega_i\;|f(\omega_s,\omega_i)|^2\times\,\mathrm{exp}\left[-\frac{(\omega_s-\omega_{s0})^2}{\sigma_f^2}\right]-    p_{++}^\mathrm{Gaus},
    \label{eq:pplusminusGaus}
\end{equation}
\begin{equation}
    p_{-+}^\mathrm{Gaus} = \int_{-\infty}^{+\infty} \mathrm{d}\omega_s \int_{-\infty}^{+\infty} \mathrm{d}\omega_i\;|f(\omega_s,\omega_i)|^2\times\,\mathrm{exp}\left[-\frac{(\omega_i-\omega_{i0})^2}{\sigma_f^2}\right]-    p_{++}^\mathrm{Gaus}
\end{equation}
and
\begin{equation}
    p_{--}^\mathrm{Gaus} = 1-p_{+-}^\mathrm{Gaus}-p_{-+}^\mathrm{Gaus}-p_{++}^\mathrm{Gaus}.
\end{equation}

If we denote the probability for the SPDC source to produce $k$ pairs of $X$-polarized signal photons and $Y$-polarized idler photons in an individual photon generation event by $\pi_{XY}(k)$, then the joint probability for $m_H$ H-polarized photons and $m_V$ V-polarized photons to enter Alice's WDM channel and $n_H$ H-polarized photons and $n_V$ V-polarized photons to enter Bob's WDM channel is equal to
 \begin{eqnarray}
        && q(m_H, m_V, n_H, n_V)    = 
         \sum_{\alpha = 0}^{\mathrm{min}[m_H, n_V]} \sum_{\beta =  0}^{\mathrm{min}[m_V, n_H]} \sum_{\gamma = 0}^{\infty} \sum_{\delta = 0}^{\infty}\,\pi_{HV} (m_H+n_V+\gamma - \alpha)\nonumber\\
        &&\times\,\pi_{VH} (m_V+n_H+\delta - \beta) \, p_{++}^{\alpha + \beta} \, p_{+-}^{m_H+m_V - \alpha - \beta} \, p_{-+}^{n_H+n_V - \alpha - \beta}\,p_{--}^{\gamma + \delta}\binom{m_H+n_V+\gamma - \alpha}{\alpha} \nonumber\\
        &&\times\,\binom{m_H+n_V+\gamma - 2\alpha}{\gamma} \binom{m_H+n_V-2\alpha}{m_H - \alpha}\binom{m_V+n_H - 2\beta}{m_V - \beta}
         \binom{m_V+n_H+\delta - 2\beta}{\delta} \nonumber\\
        &&\times\, 
        \binom{m_V+n_H+\delta - \beta}{\beta} .
\end{eqnarray}   
Finally, the probability $p (i_H, i_V, j_H, j_V)$ can be calculated as
\begin{eqnarray}
        \label{eq:clickprobgeneral}
       && p (i_H, i_V, j_H, j_V) =\sum_{m_H=i_H}^{\infty} \sum_{m_V=i_V}^{\infty}
        \sum_{n_H=j_H}^{\infty}
        \sum_{n_V=j_V}^{\infty} \binom{m_H}{i_H} \binom{m_V}{i_V} \binom{n_H}{j_H} \binom{n_V}{j_V}\\
        && \times\,q(m_H, m_V, n_H, n_V)\,T_A ^{i_H+i_V}(1-T_A) ^{m_H+m_V - i_H - i_V}\,T_B ^{j_H+j_V} \, (1-T_B)^{n_H+n_V-j_H-j_V}.\nonumber
\end{eqnarray}

By inserting the relevant formulas for $p_\mathrm{acc}$ and $Q$, derived above, with $p (i_H, i_V, j_H, j_V)$ given by Eq.\,(\ref{eq:clickprobgeneral}), into Eq.\,(\ref{eq:keyratemain}) one can estimate lower bound on the key generation rate produced from a pair of WDM channels, centered at the angular frequency detunings $\left\{\omega_{s0},\omega_{i0}\right\}$. The overall key rate, generated from all of the channel pairs utilized by Alice and Bob, can be calculated as the sum of individual key rates obtained for the set of angular frequencies provided by Eq.\,(\ref{eq:omegasetnegative}) or Eq.\,(\ref{eq:omegasetpositive}), depending on the type of spectral correlation generated between the SPDC photons. In the next section we present the results of such calculations. 

\section{Numerical simulation}
\label{Sec:Results}

\subsection{Key generation without WDM modules}
\label{Sec:NoWDM}

\begin{figure}[tb]
\centering
\includegraphics[width=\linewidth]{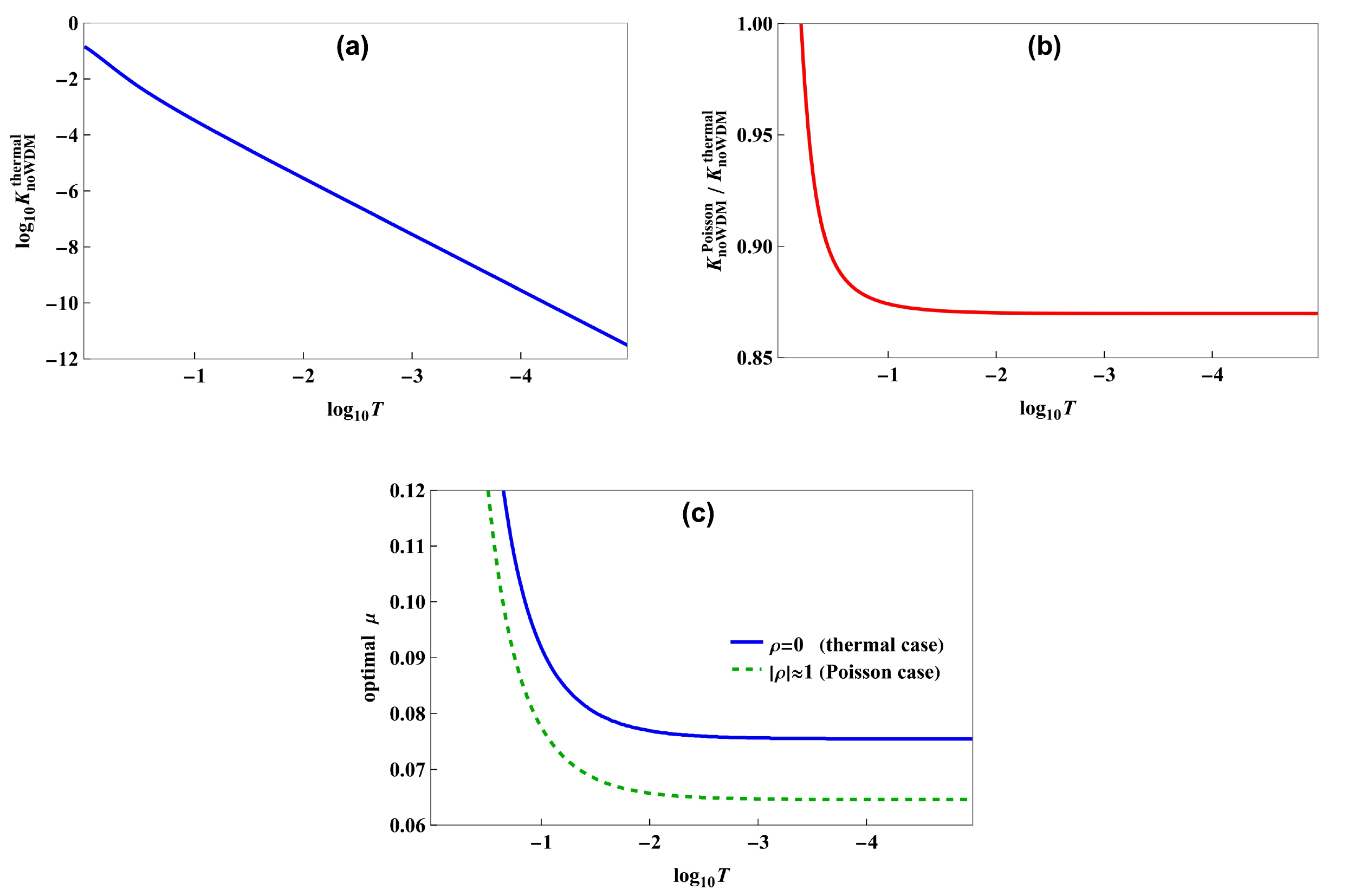}
\caption{(a) Lower bound for the key generation rate that can be obtained by Alice and Bob with the use of the SPDC source producing spectrally uncorrelated photons, $\rho=0$, in traditional QKD setup without WDM modules, plotted as a function of the transmittance of the symmetric channels connecting the source with the trusted parties, $T$. (b) The ratio between the key generation rate that can be obtained by Alice and Bob with the use of the SPDC source producing photons with strong spectral correlation, $|\rho|\approx1$, in traditional QKD setup without WDM modules, and the key generation rate illustrated in panel (a). The calculated key generation rates are optimized over the mean number of photons produced by the SPDC source in a single polarization mode per one attempt, $\mu$. (c) The optimal values of $\mu$ for the cases of spectrally uncorrelated (blue solid line) and strongly correlated (green dashed line) photons.}
\label{fig:noWDM}
\end{figure}

Before focusing on the setup configuration presented in Fig.\,\ref{fig:setup} let us estimate the key generation rate that can be obtained by the trusted parties when utilizing the traditional scheme, without WDM modules. In this case the formula (\ref{eq:clickprobgeneral}) simplifies to
\begin{eqnarray}
       && p_\mathrm{no WDM}(i_H, i_V, j_H, j_V) =\sum_{m=\mathrm{max}\left\{i_H,j_V\right\}}^{\infty} \sum_{n=\mathrm{max}\left\{i_V,j_H\right\}}^{\infty}  q_\mathrm{no WDM}(m,n)\,T_A ^{i_H+i_V}\nonumber\\
        && \times\,(1-T_A) ^{m+n - i_H - i_V}\,T_B ^{j_H+j_V} \, (1-T_B)^{m+n-j_H-j_V},
        \label{eq:clickprobnoWDM}
\end{eqnarray}
where
\begin{equation}
    q_\mathrm{no WDM}(m,n)=\pi_{HV}(m)\,\pi_{VH} (n)
\end{equation}
is the joint probability for the source to produce $m$ pairs of $H$-polarized signal and $V$-polarized idler photons and $n$ pairs of $V$-polarized signal and $H$-polarized idler photons from a single pump pulse.

The key generation rate depends on the statistics of photon pairs produced by the SPDC source. For single-spectral-mode SPDC process, which takes place when no spectral correlation between the signal and idler photons can be observed, \emph{i.e.} $\rho=0$, the number of photon pairs generated in each of the two polarization modes follows thermal probability distribution \cite{Mauerer2009}. The lower bound for the key generation rate, denoted by $K_\mathrm{no WDM}^\mathrm{thermal}$, produced in this case, optimized over the power of the source, which is proportional to the parameter $\mu$, calculated as a function of the transmittance of the channels connecting the source with the trusted parties, is presented in Fig.\,\ref{fig:noWDM}\,(a). The plot is only slightly deviated from the linear function, as could be expected from the log-log type of key generation rate plot in the noiseless case. The deviation can be observed only for relatively high values of the channel transmittance and its origin stems from the non-negligible number of multi-click events that are registered in this scenario.

On the other hand, in the limit of very strong spectral correlation, $|\rho|\approx1$, the number of contributing spectral modes approaches infinity and the photon pair statistics becomes Poissonian \cite{Mauerer2009}. The comparison between the key generation rates that can be obtained by Alice and Bob in the Poissonian ($K_\mathrm{no WDM}^\mathrm{Poisson}$) and thermal ($K_\mathrm{no WDM}^\mathrm{thermal}$) cases is shown in Fig.\,\ref{fig:noWDM}\,(b). Unless the transmittance of the channels is very high, the rate calculated for the thermal statistics of the emitted photon pairs is slightly higher. For $T\rightarrow0$ the difference asymptotically approaches $13\%$. To understand the reason for such outcome let us consider the events when the source produces two photon pairs in total. In this case the error in the raw key is generated only if one $HV$ pair and one $VH$ pair were emitted, and the trusted parties detect photons from two different pairs. If two $HV$ or two $VH$ pairs were produced the error would not appear regardless of which photons are ultimately detected. In the thermal case the joint probability for one $HV$ pair and one $VH$ pair to be emitted is equal to $\mu^2/(\mu+1)^4$ and is the same as the probability for two $HV$ or two $VH$ pairs to be produced. On the other hand in the Poissonian case the probability for one $HV$ pair and one $VH$ pair to be emitted, equal to $\mu^2e^{-2\mu}$, is two times larger than the generation probabilities for either two $HV$ or two $VH$ pairs. After performing similar analysis for the scenarios of more than two photon pairs being created by the SPDC source from a single pump pulse, one can arrive at the conclusion that for a given value of $\mu$ the higher QBER should be expected in the case when the number of photon pairs produced in a single polarization mode follows Poissonian statistics. Thus, the optimal value of $\mu$ for the thermal case turns out to be slightly higher than in the Poissonian case, as can be seen in Fig.\,\ref{fig:noWDM}\,(c), leading to larger key generation rate.

\subsection{Key generation with WDM modules -- negative spectral correlation}
\label{Sec:WDMnegative}

Taking the results presented in Fig.\,\ref{fig:noWDM} into account, it is reasonable to adopt $K_\mathrm{no WDM}^\mathrm{thermal}$ as the reference to which we compare the key generation rate obtained by the trusted parties when using WDM modules in their QKD setup, as pictured in Fig.\,\ref{fig:setup}. For this purpose let us define the gain function
\begin{equation}
    G=K_\mathrm{WDM}^\mathrm{total}/K_\mathrm{no WDM}^\mathrm{thermal},
    \label{eq:gain}
\end{equation}
where $K_\mathrm{WDM}^\mathrm{total}$ is the sum of all the key rates estimated with the use of the formulas presented in Sec.\,\ref{Sec:Theory} for each pair of WDM channels utilized by Alice and Bob for the separate key generation process. All the calculations of $K_\mathrm{WDM}^\mathrm{total}$, the results of which are presented below, were performed with the assumption that the statistics of photon pairs emitted by the SPDC source can be approximated with the Poisson probability distribution. This simplification is reasonable when the spectral correlation between the signal and idler photons is relatively strong. While it changes the obtained key rates slightly, this effect is insignificant to the derived conclusions (see the Appendix for more detailed discussion).

\begin{figure}[tb]
\centering
\includegraphics[width=\linewidth]{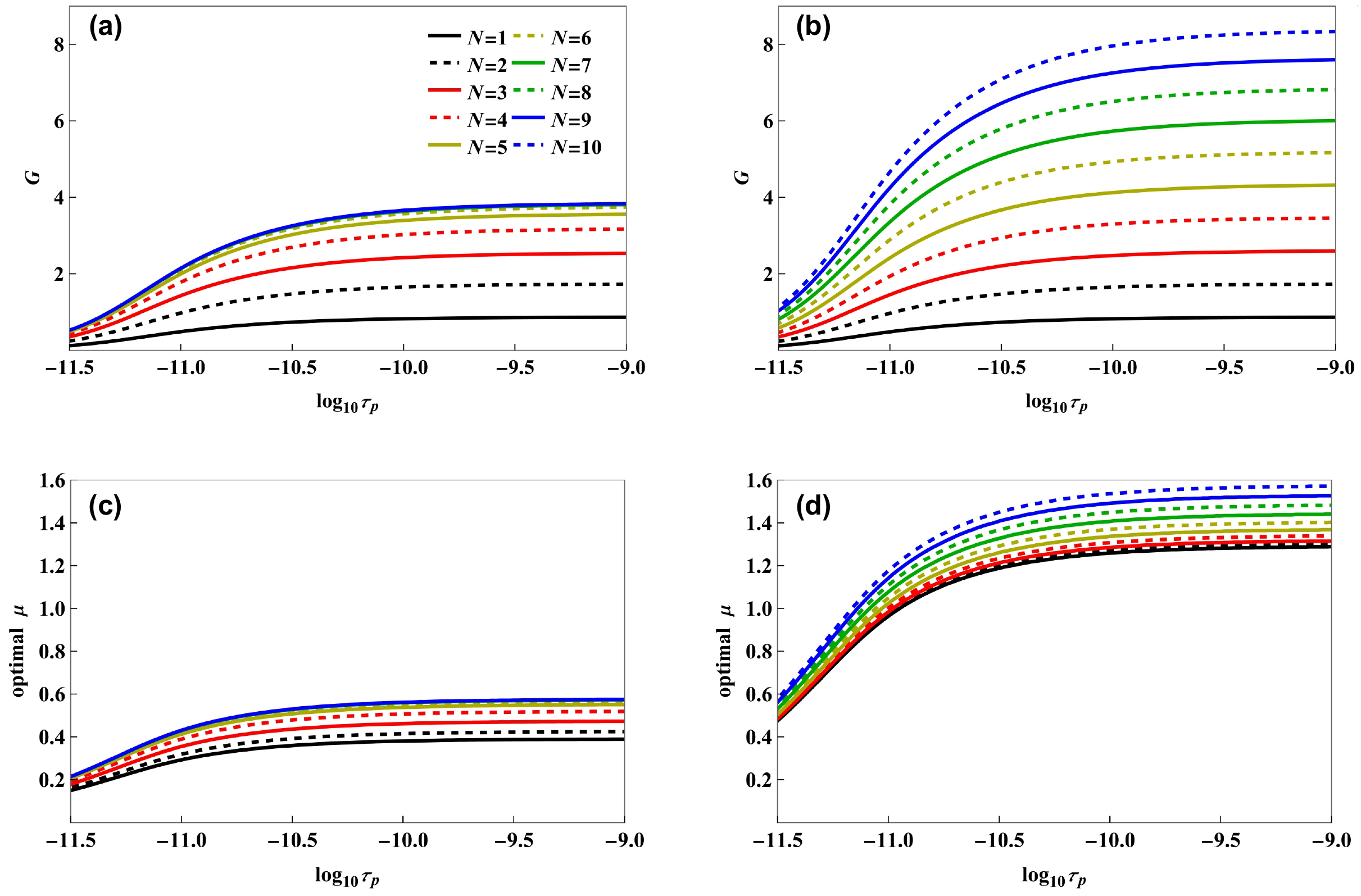}
\caption{The gain function, defined with Eq.\,(\ref{eq:gain}), plotted against the pump laser pulse duration, $\tau_p$, for the EPMF width equal to (a) $\sigma_{cr}=3\,\mathrm{THz}$ and (b) $\sigma_{cr}=10\,\mathrm{THz}$. The plots are maximized over the mean number of photon pairs emitted by the SPDC source in a single polarization mode $\mu$ and the optimal values of this parameter are shown in panels (c) and (d), respectively. Different colors and styles of all the lines correspond to various numbers of WDM channel pairs, $N$, utilized by Alice and Bob, as shown in the legend in panel (a). The transmission profiles of WDM channels is assumed to be rectangular and their central angular frequencies are defined according to Eq.\,(\ref{eq:omegasetnegative}). All the results are calculated for the transmittance of the channels connecting the source with the trusted parties equal to $T=10^{-3}$.}
\label{fig:KeyRateVsTaupNeg}
\end{figure}

\begin{figure}[tb]
\centering
\includegraphics[width=\linewidth]{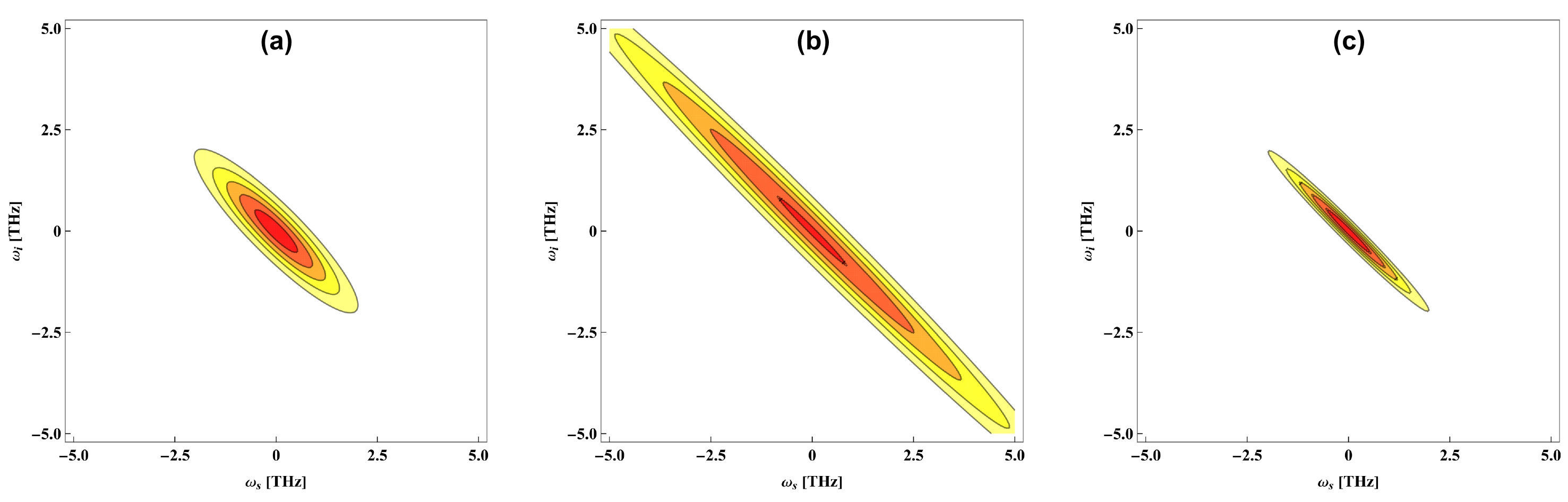}
\caption{Spectral biphoton wavefunction $f(\omega_s,\omega_i)$, given by Eq.\,(\ref{eq:wavefunction}), plotted for (a) $\sigma_{cr}=3\,\mathrm{THz}$ and $\tau_p=3\,\mathrm{ps}$, (b) $\sigma_{cr}=10\,\mathrm{THz}$ and $\tau_p=3\,\mathrm{ps}$, (c) $\sigma_{cr}=3\,\mathrm{THz}$ and $\tau_p=8\,\mathrm{ps}$.}
\label{fig:SpectralBiphotonWavefunction}
\end{figure}

We start the analysis from the scenario of the negative spectral correlation between the signal and idler photons produced by the SPDC source, in which case it is reasonable to establish the pairs of WDM channels according to the formula (\ref{eq:omegasetnegative}). In Fig.\,\ref{fig:KeyRateVsTaupNeg}\,(a) we present the dependence of the gain function on the pump laser pulse duration for various numbers of WDM channel pairs utilized to produce the key, calculated for $\sigma_{cr}=3\,\mathrm{THz}$ and optimized over the parameter $\mu$. While the plots were made for $T=10^{-3}$, the results are approximately independent from the transmittance of the channel as long as $T<10^{-1}$. As shown in Fig.\,\ref{fig:KeyRateVsTaupNeg}\,(a), regardless of the number of WDM channel pairs utilized by Alice and Bob, the gain $G$ grows with increasing pump laser pulse duration, reaching an asymptotic value for $\tau_p\rightarrow\infty$. When $N$ is small, this value is equal to $NK_\mathrm{no WDM}^\mathrm{Poisson}/K_\mathrm{no WDM}^\mathrm{thermal}$. However, when $N>6$ the $G$ function stops increasing further with $N$. The reason for this behaviour is the assumed value of the EPMF width $\sigma_{cr}$, which is the primary parameter responsible for the width of the spectral biphoton wavefunction $f(\omega_s,\omega_i)$ in the diagonal direction on the $\{\omega_s,\omega_i\}$ grid. This can be seen by comparing the panels (a) and (b) in Fig.\,\ref{fig:SpectralBiphotonWavefunction}. As it turns out, for $\sigma_{cr}=3\,\mathrm{THz}$ the values assumed by $f(\omega_s,\omega_i)$ for channel pairs with numbers $n<-5$ or $n>5$ are so small that there is no significant contribution to the overall key rate coming from them. Therefore, extending the number of WDM channel pairs utilized by the trusted parties to more than six does not improve $K_\mathrm{WDM}^\mathrm{total}$ anymore. In order to better the situation one has to increase $\sigma_{cr}$. This is confirmed in Fig.\,\ref{fig:KeyRateVsTaupNeg}\,(b), where $\sigma_{cr}=10\,\mathrm{THz}$ and one can observe steady growth of the gain function with $N$ at least up to $N=10$.

Broader biphoton wavefunction in the diagonal direction means also that the probability for signal and idler photons to enter a given pair of WDM channels becomes smaller. Therefore, the optimal values of $\mu$, maximizing the overall key generation rate, are significantly higher for $\sigma_{cr}=10\,\mathrm{THz}$ than for $\sigma_{cr}=3\,\mathrm{THz}$. This can be confirmed by comparing the panels (c) and (d) of Fig.\,\ref{fig:KeyRateVsTaupNeg}. There, one can also observe an asymptotic growth of optimal $\mu$ with $\tau_p\rightarrow\infty$, regardless of the EPMF width or the number of WDM channel pairs utilized by Alice and Bob. This growth can be explained by the fact that pump laser pulse duration is the primary factor governing the width of the spectral biphoton wavefunction in the antidiagonal direction, as can be seen by comparing the panels (a) and (c) in Fig.\,\ref{fig:SpectralBiphotonWavefunction}. The longer $\tau_p$ is, the narrower $f(\omega_s,\omega_i)$ becomes, lowering the probabilities $p_{+-}$ and $p_{-+}$, given by the formulas (\ref{eq:pplusminusrect}) and (\ref{eq:pminusplusrect}) respectively, that contribute to the QBER. Therefore, while the pump laser pulse duration does not influnece the number of channel pairs from which Alice and Bob can get significant contribution to the overall key generation rate, its value becomes important for the key rate that can be obtained from every single pair.

\begin{figure}[tb]
\centering
\includegraphics[width=0.6\linewidth]{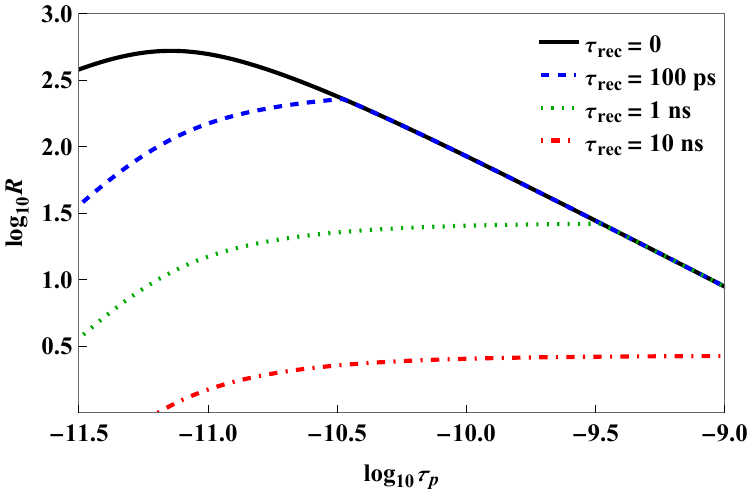}
\caption{The lower bound for the number of bits of the secure key per unit of time, $R$, that can be obtained by Alice and Bob from the central WDM channels pair (\emph{i.e.} the one with $n=0$ in Eq.\,(\ref{eq:omegasetnegative})) with rectangular transmission profile, plotted as a function of pump laser pulse duration, $\tau_p$, assuming that the repetition rate of the source is given by the formula (\ref{eq:reprate}). Four different values of the recovery time $\tau_\mathrm{rec}$ are considered, as indicated in the legend. The plots are made for the EPMF width $\sigma_{cr}=10\,\mathrm{THz}$ and the transmittance of the channels connecting the source with the trusted parties equal to $T=10^{-3}$.}
\label{fig:KeyRatePerSecond}
\end{figure}

From the above analysis it would appear that regardless of the EPMF width it is beneficial for the trusted parties to choose $\tau_p$ as long as possible. However, in practice it is more important to maximize the key rate that can be obtained per unit of time, $R$, than the key rate per one use of the source, given by the formula (\ref{eq:keyratemain}). The relationship between the two quantities is
\begin{equation}
    R=f_rK,
    \label{eq:keypersecond}
\end{equation}
where $f_r$ is the repetition rate of the source. In realistic scenario $f_r$ is primarily limited by the longest recovery time of the utilized setup elements $\tau_\mathrm{rec}$ (\emph{e.g.} the time needed by the source of photons to produce another signal or the dead time of the detectors), which is generally independent from the signal properties. However, if the pump pulses were very long, the main limitation on $f_r$ would come from the requirement for proper separation between the subsequent signals emitted by the source. In Fig.\,\ref{fig:KeyRatePerSecond} we plot the key generation rate per unit of time, optimized over $\mu$, calculated as a function of the pump laser pulse duration for the case when $\sigma_{cr}=10\,\mathrm{THz}$. For the purpose of this calculation we assumed that reasonable separation between the pump laser pulses is equal to $3\tau_p$ and therefore
\begin{equation}
    f_r=\mathrm{min}\left\{1/\tau_\mathrm{rec},1/(3\tau_p)\right\}.
    \label{eq:reprate}
\end{equation} 
The plots were made for four different values of $\tau_\mathrm{rec}$. While the figure illustrates only the case of a single WDM channel pair utilized by Alice and Bob, the general behaviour of the calculated function is independent of $N$. The conclusion from Fig.\,\ref{fig:KeyRatePerSecond} is that as long as $\tau_\mathrm{rec}>10\,\mathrm{ps}$ (which would be extremely hard to break with the use of currently available technology) the optimal value of pump laser pulse duration, maximizing the key generation rate per unit of time, would be equal to $\tau_p^\mathrm{opt}=\tau_\mathrm{rec}/3$.

\begin{figure}[tb]
\centering
\includegraphics[width=\linewidth]{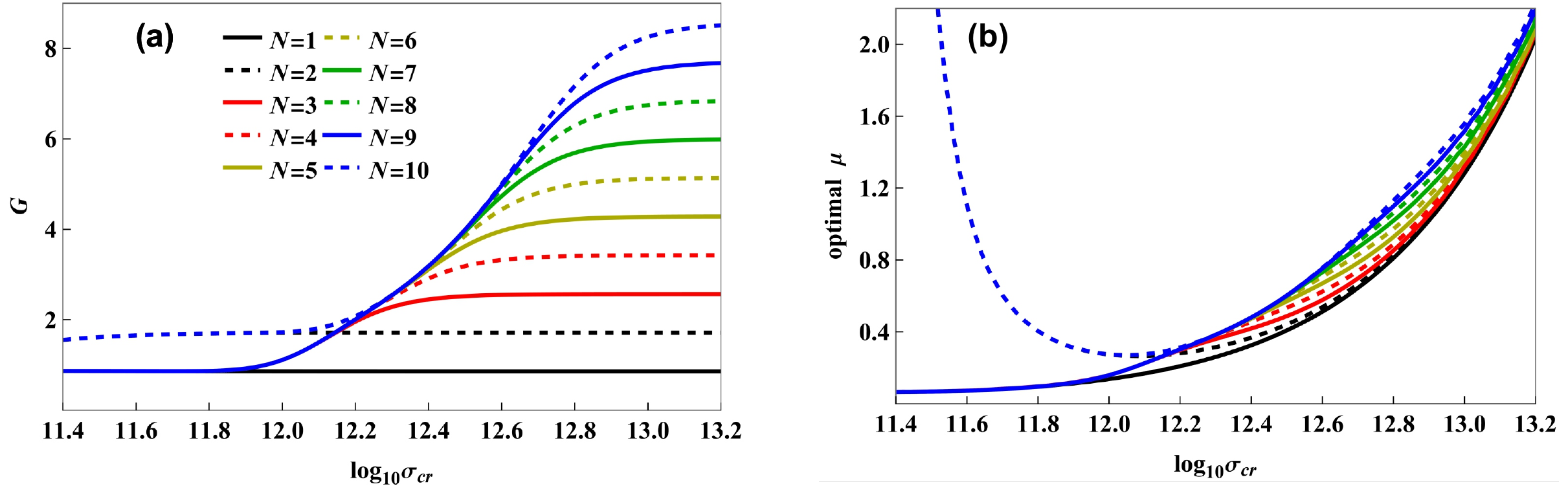}
\caption{(a) The gain function, defined with Eq.\,(\ref{eq:gain}), plotted against the EPMF width, $\sigma_{cr}$, for the pump laser pulse duration equal to $\tau_{p}=333\,\mathrm{ps}$. The plots are maximized over the mean number of photon pairs emitted by the SPDC source in a single polarization mode, $\mu$. Different colors and styles of the lines correspond to various numbers of WDM channel pairs, $N$, utilized by Alice and Bob, as shown in the legend. The transmission profiles of WDM channels are assumed to be rectangular and their central angular frequencies are defined according to Eq.\,(\ref{eq:omegasetnegative}). All the results are calculated for the transmittance of the channels connecting the source with the trusted parties equal to $T=10^{-3}$. (b) The optimal values of $\mu$, corresponding to the results illustrated in panel (a).}
\label{fig:KeyRateVsEPMFNeg}
\end{figure}

In realistic situation it is reasonable to expect the setup recovery time of the order of one nanosecond. Therefore, in Fig.\,\ref{fig:KeyRateVsEPMFNeg}\,(a) we plot the gain function optimized over $\mu$ against the EPMF width with the assumption that $\tau_p^\mathrm{opt}\approx333\,\mathrm{ps}$. As already concluded from the comparison between the panels (a) and (b) of Fig.\,\ref{fig:KeyRateVsTaupNeg}, the higher $\sigma_{cr}$ becomes, the more WDM channel pairs can contribute to the overall key generation rate, which could potentially grow to infinity for $\sigma_{cr}\rightarrow\infty$. However, in practice this paramater of the SPDC source is always limited, depending on the type of utilized nonlinear material. For example in the case of bulk BBO crystal the values of $\sigma_{cr}$ between $10^{11}\,\mathrm{Hz}$ and $10^{13}\,\mathrm{Hz}$ were shown to be realistically achievable \cite{Lasota2020}. Fig.\,\ref{fig:KeyRateVsEPMFNeg}\,(a) can be also used to assess the number of WDM channel pairs that can be effectively employed to increase the key generation rate in situations when the SPDC source is already decided and the value of the EMPF width is fixed.

The optimal mean number of photon pairs produced by the source in a single polarization mode, corresponding to the results presented in Fig.\,\ref{fig:KeyRateVsEPMFNeg}\,(a), is illustrated in the panel (b) of the same figure. With $\sigma_{cr}\rightarrow\infty$ this quantity also grows to infinity. However, for even number of WDM channel pairs utilized by Alice and Bob one can observe $\mu\rightarrow\infty$ also when the value of $\sigma_{cr}$ is very small. Similar behavior cannot be seen when $N$ is odd. The reason for this difference stems from our assumption regarding the WDM channel positions, explained in details in Sec.\ref{Sec:Assumptions}. When $\sigma_{cr}<1\mathrm{THz}$ the spectral biphoton wavefunction $f(\omega_{s},\omega_{i})$ becomes so narrow in the diagonal direction that the probability for the SPDC photon pair to enter any pair of WDM channels beside the one with $n=0$ is close to zero. However, this pair of channels is part of the WDM grid only in the odd $N$ case. Therefore, when $N$ is even the power of the pump laser has to be significantly higher to produce decent overall key rate.

\subsection{Key generation with WDM modules -- positive spectral correlation}
\label{Sec:WDMpositive}

\begin{figure}[tb]
\centering
\includegraphics[width=\linewidth]{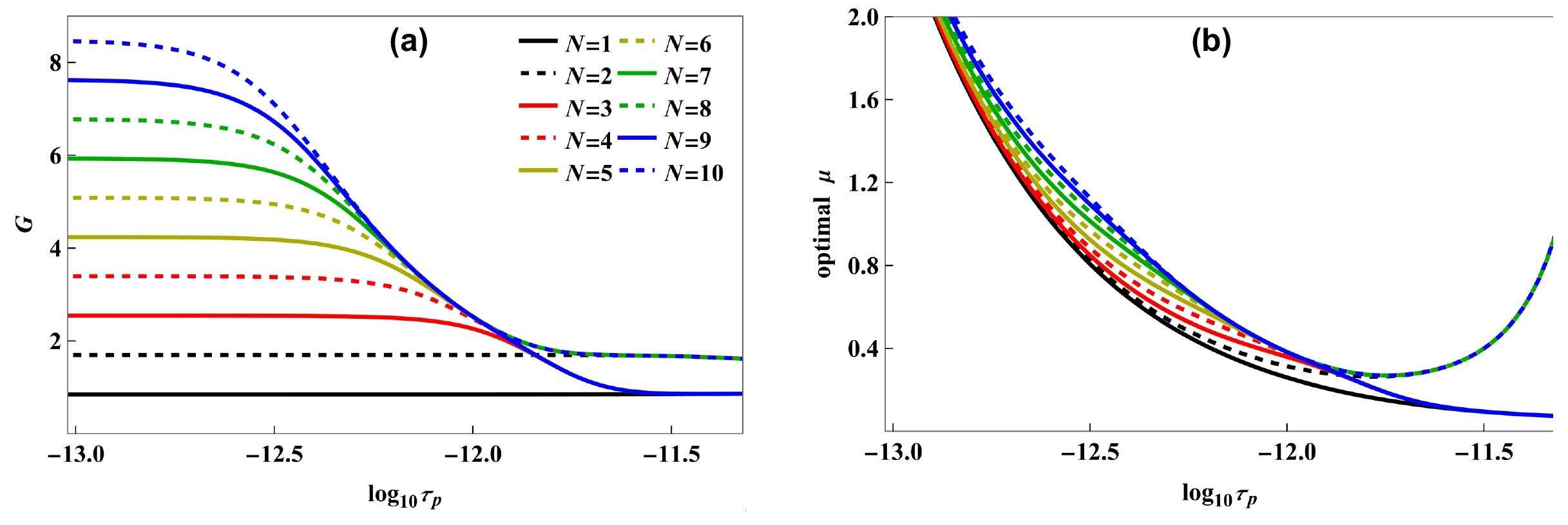}
\caption{(a) The gain function, defined with Eq.\,(\ref{eq:gain}), plotted against the pump laser pulse duration, $\tau_p$, for the EPMF width equal to $\sigma_{cr}=10\,\mathrm{GHz}$. Different colors and styles of all the lines correspond to various numbers of WDM channel pairs, $N$, utilized by Alice and Bob, as shown in the legend. The transmission profiles of WDM channels are assumed to be rectangular and their central angular frequencies are defined according to Eq.\,(\ref{eq:omegasetpositive}). The results are calculated for the transmittance of the channels connecting the source with the trusted parties equal to $T=10^{-3}$. The plots are maximized over the mean number of photon pairs emitted by the SPDC source in a single polarization mode, $\mu$. (b) Optimal values of $\mu$ corresponding to the results shown in panel (a).}
\label{fig:KeyRateVsTaupPos}
\end{figure}

Utilizing WDM modules to increase $K_\mathrm{WDM}^\mathrm{total}$ is also possible when the SPDC source generates photons with strong positive spectral correlation. In this situation the corresponding WDM channel pairs should be defined according to the formula (\ref{eq:omegasetpositive}). As illustrated in Fig.\,\ref{fig:spectrum}\,(b) they are placed on the anti-diagonal of the $\{\omega_s,\omega_i\}$ grid. Therefore, the effects of changing the values of the source parameters, $\sigma_{cr}$ and $\tau_p$, on the overall key rate are reversed in comparison to the case of negative spectral correlation, analyzed in Sec.\,\ref{Sec:WDMnegative}. 

When the positively correlated photons are produced, the modification of the pump laser pulse duration is responsible for adjusting the number of WDM channel pairs, from which the trusted parties can be able to obtain non-negligible contribution to the overall key. This effect is presented in Fig.\,\ref{fig:KeyRateVsTaupPos}\,(a) where we plotted the gain $G$ as a function of $\tau_p$ for $\sigma_{cr}=10\,\mathrm{GHz}$ and for up to ten WDM channel pairs used by Alice and Bob. The observed behavior resembles the $G(\sigma_{cr})$ dependence for the case of negative spectral correlation, shown previously in Fig.\,\ref{fig:KeyRateVsEPMFNeg}\,(a). Also the corresponding plot of the optimal mean number of photon pairs generated in single polarization mode as a function of the pump laser pulse duration, illustrated in Fig.\,\ref{fig:KeyRateVsTaupPos}\,(b), exhibits evident similarity to Fig.\,\ref{fig:KeyRateVsEPMFNeg}\,(b). From Fig.\,\ref{fig:KeyRateVsTaupPos}\,(a) one can conclude that the shorter the pump laser pulse duration is, the more WDM channel pairs can contribute to the overall key generation rate and it is theoretically possible to increase this quantity to arbitrary values when $\tau_p\rightarrow0$. However, similarly as the realistic values of the EPMF width bound the achievable key rate in the case of negative spectral correlation, limited ability to produce ultrashort laser pulses does it in the positive-correlation scenario.

\begin{figure}[tb]
\centering
\includegraphics[width=\linewidth]{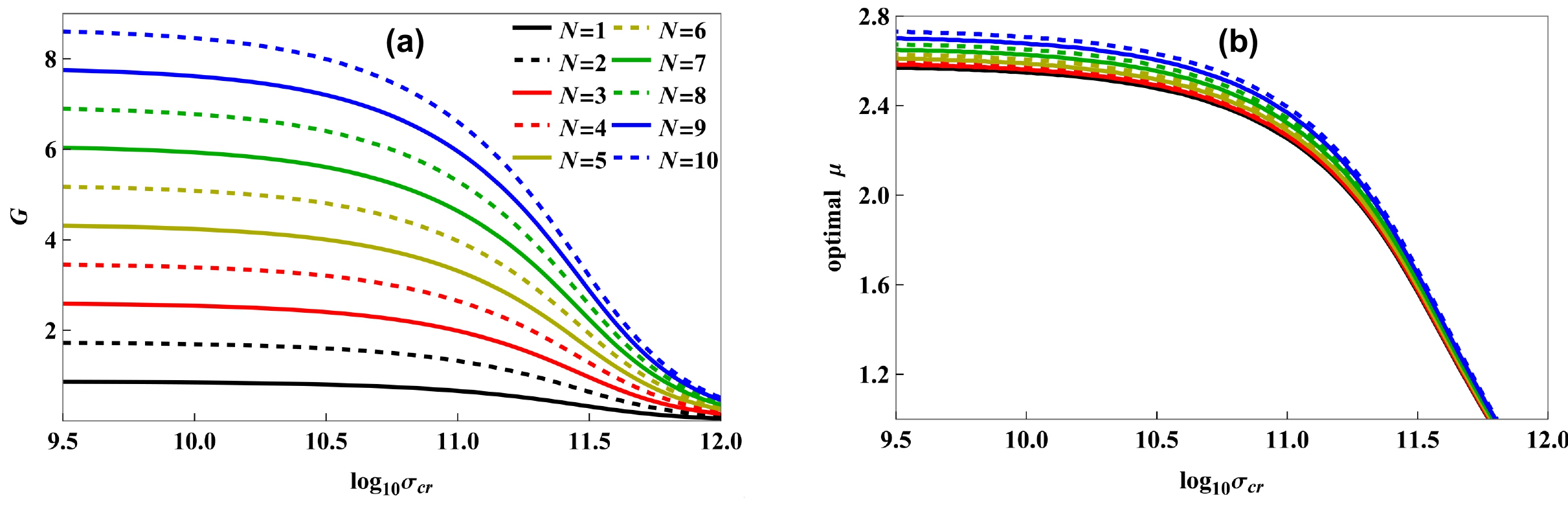}
\caption{(a) The gain function, defined with Eq.\,(\ref{eq:gain}), plotted against the EPMF width, $\sigma_{cr}$, for the pump laser pulse duration equal to $\tau_{p}=100\,\mathrm{fs}$. The plots are maximized over the mean number of photon pairs emitted by the SPDC source in a single polarization mode, $\mu$. Different colors and styles of the lines correspond to various numbers of WDM channel pairs, $N$, utilized by Alice and Bob, as shown in the legend. The transmission profiles of WDM channels are assumed to be rectangular and their central angular frequencies are defined according to Eq.\,(\ref{eq:omegasetpositive}). All the results are calculated for the transmittance of the channels connecting the source with the trusted parties equal to $T=10^{-3}$. (b) The optimal values of $\mu$, corresponding to the results illustrated in panel (a).}
\label{fig:KeyRateVsEPMFPos}
\end{figure}

While in the case of positive spectral correlation the number of WDM channel pairs contributing to the overall key turns out to be almost independent of the EPMF width, $\sigma_{cr}$ is the main parameter influencing the secret key rate that can be distilled from a single pair of channels. This relationship is investigated in Fig.\,\ref{fig:KeyRateVsEPMFPos}\,(a), where the function $G(\sigma_{cr})$ is plotted for the pump laser pulse duration $\tau_p=100\,\mathrm{fs}$. When $\sigma_{cr}\rightarrow0$ the gain approximately reaches the value of $NK_\mathrm{no WDM}^\mathrm{Poisson}/K_\mathrm{no WDM}^\mathrm{thermal}$. On the other hand, for $\sigma_{cr}>100\,\mathrm{GHz}$ one can see evident decrease of the presented function. 

\subsection{Influence of WDM channels' properties on the overall key generation rate}
\label{Sec:WDMproperties}

All the plots shown in Secs.\,\ref{Sec:WDMnegative}--\ref{Sec:WDMpositive} were produced with a set of specific assumptions on the properties of individual channels provided by the WDM modules utilized by the trusted parties. In particular, we considered WDM channels with rectangular transmission profiles, given by Eq.(\ref{eq:rectchannel}), with the channel width $\Delta\omega=2\pi\times50\,\mathrm{GHz}$ and the separation between the neighboring channels equal to $\omega_\mathrm{sep}=2\pi\times100\,\mathrm{GHz}$. 
To make our analysis more general in this subsection we investigate how modifying the above assumptions can influence the results presented before.

\begin{figure}[tb]
\centering
\includegraphics[width=\linewidth]{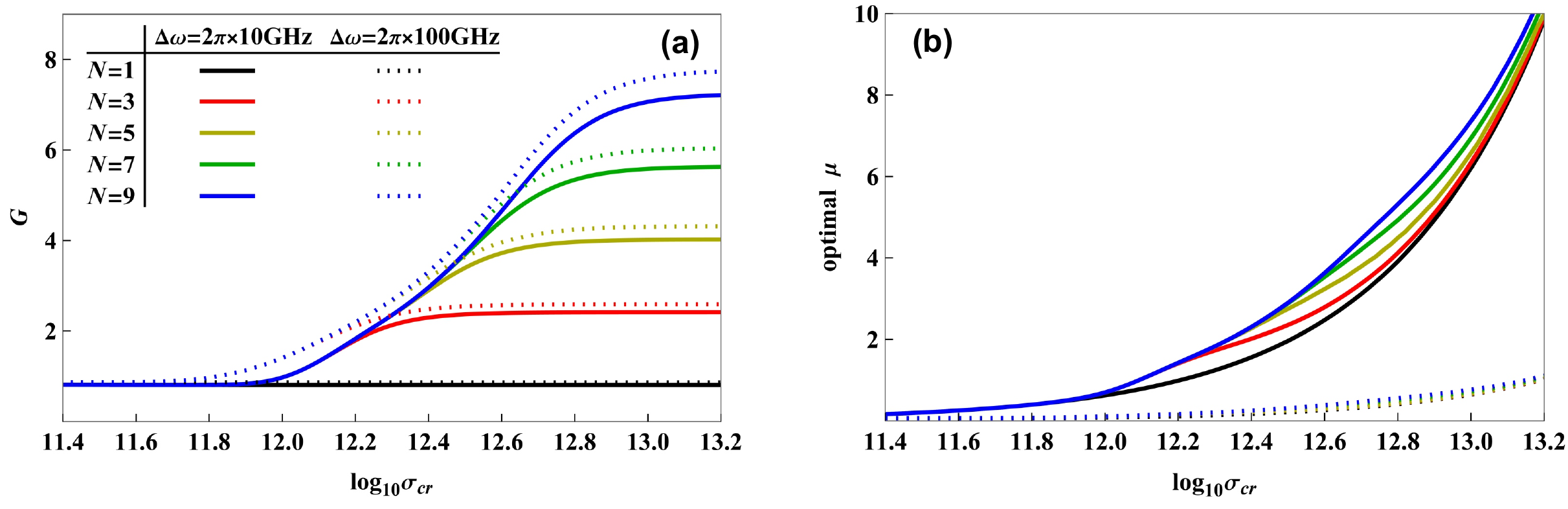}
\caption{(a) The gain function, defined with Eq.\,(\ref{eq:gain}), plotted against the EPMF width, $\sigma_{cr}$, for the pump laser pulse duration equal to $\tau_{p}=333\,\mathrm{ps}$. The plots are maximized over the mean number of photon pairs emitted by the SPDC source in a single polarization mode, $\mu$. Different colors and styles of the lines correspond to various numbers of WDM channel pairs, $N$, utilized by Alice and Bob, and the widths of the WDM channels, respectively, as indicated in the legend. The transmission profiles of the WDM channels are assumed to be rectangular and their central angular frequencies are defined according to Eq.\,(\ref{eq:omegasetnegative}). The separation between the neighboring channels is equal to $\omega_\mathrm{sep}=2\pi\times100\,\mathrm{GHz}$. All the results are calculated for the transmittance of the channels connecting the source with the trusted parties equal to $T=10^{-3}$. (b) The optimal values of $\mu$, corresponding to the results illustrated in panel (a).}
\label{fig:KeyRateWDMwidth}
\end{figure}

Let us first focus on the channel width. In Fig.\ref{fig:KeyRateWDMwidth}\,(a) we compare the gain, $G$, plotted as a function of the EPMF width, calculated for two different values of $\Delta\omega$. The case of $\Delta\omega=2\pi\times100\,\mathrm{GHz}$ can be treated as an ideal situation, in which the channels are completely separated from each other, although there is no gap between them. On the other hand WDM module with the channel width of $\Delta\omega=2\pi\times10\,\mathrm{GHz}$ and separation of $\omega_\mathrm{sep}=2\pi\times100\,\mathrm{GHz}$ would be quite ineffective, as most of the signal and idler photons would be blocked by it. To make Fig.\ref{fig:KeyRateWDMwidth}\,(a) easier to read only the results for odd number of channel pairs utilized by Alice and Bob are shown. The outcomes for even $N$ are completely analogous.

The difference between the values of $G$ that can be obtained for $\Delta\omega=2\pi\times100\,\mathrm{GHz}$ and $\Delta\omega=2\pi\times10\,\mathrm{GHz}$ turns out to be relatively small, because larger loss of photons, observed in the latter case, can be neutralized by increasing the mean number of photons produced by the SPDC source in a single polarization mode per pulse. This can be clearly seen in Fig.\ref{fig:KeyRateWDMwidth}\,(b), where we plot the optimal $\mu$ corresponding to the results shown in panel (a). Still, the maximal value of the gain function turns out to be slightly higher for $\Delta\omega=2\pi\times100\,\mathrm{GHz}$. There are two reasons for this outcome. Firstly, the relative difference between the values of the probabilities $p_{++}^\mathrm{rect}$, calculated for individual channel pairs, grow when the channels become narrower. This means that the values of $\mu$, optimizing the key rate provided by each of these pairs, are more imbalanced, which negatively affects the optimization capability of the overall system. Furthermore, when the channels are narrower, the relative width of the spectral biphoton wavefunction for a given value of $\tau_p$ increases. It negatively affects the QBER and slightly decreases the key generation rate obtainable from each individual WDM channel pair.

While the previously assumed separation between the neighboring WDM channels is standard for classical communication networks, WDM modules with other values of $\omega_\mathrm{sep}$ can be also utilized. Therefore, it is useful to investigate how the overall key generation rate provided by the QKD system with $N$ WDM channel pairs would change if the channel separation is modified from $\omega_\mathrm{sep}$ to $A\omega_\mathrm{sep}$, where $A$ is an arbitrary positive number. This task can be easily resolved by realizing that the probabilities $p_{++}^\mathrm{rect}$, $p_{+-}^\mathrm{rect}$, $p_{-+}^\mathrm{rect}$ and $p_{--}^\mathrm{rect}$ are invariant under the following transformation of the relevant variables:
\begin{equation}
\left\{\begin{array}{lll}
\omega_\mathrm{sep}&\rightarrow&A\omega_\mathrm{sep}\\
\Delta\omega&\rightarrow&A\Delta\omega\\
\sigma_{cr}&\rightarrow&A\sigma_{cr}\\
\tau_p&\rightarrow&\tau_p/A
  \end{array}\right.\,.
  \label{eq:rescaling}
\end{equation}
Thus, the overall key generation rate for the case of WDM channel separation equal to $A\omega_\mathrm{sep}$ can be found by considering the standard value of $\omega_\mathrm{sep}=2\pi\times100\,\mathrm{GHz}$ with channel width and SPDC source properly adjusted, according to Eq.\,(\ref{eq:rescaling}).

\begin{figure}[tb]
\centering
\includegraphics[width=\linewidth]{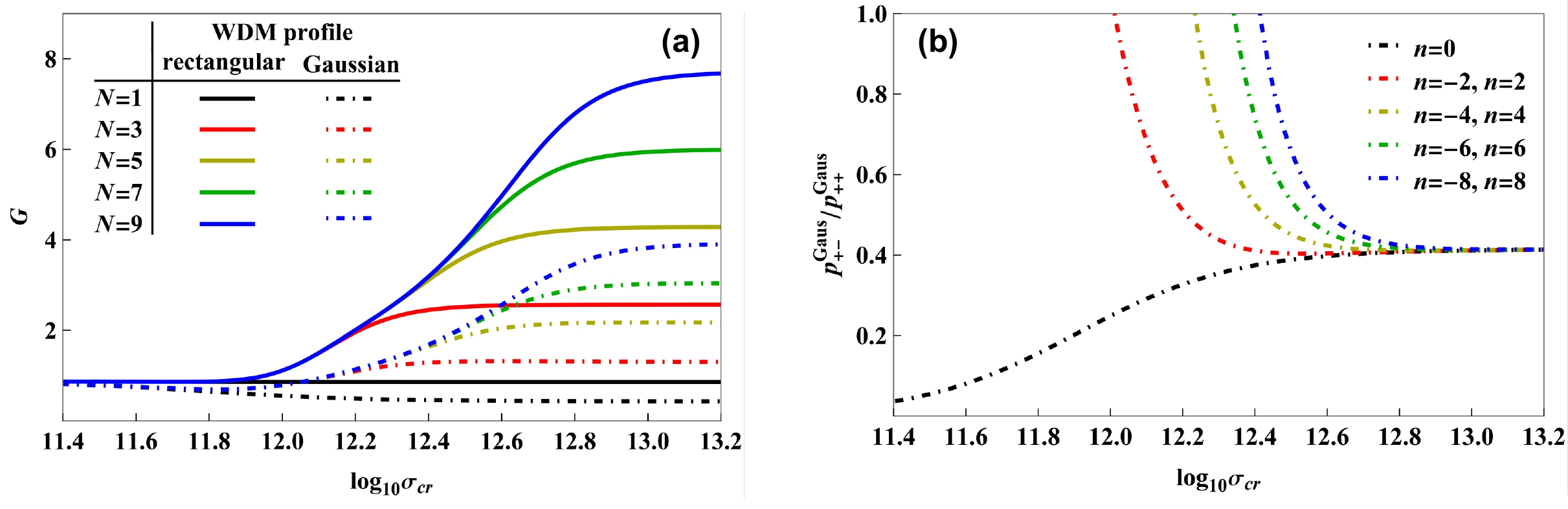}
\caption{(a) The gain function, defined with Eq.\,(\ref{eq:gain}), plotted against the EPMF width, $\sigma_{cr}$, for the pump laser pulse duration equal to $\tau_{p}=333\,\mathrm{ps}$. The plots are maximized over the mean number of photon pairs emitted by the SPDC source in a single polarization mode, $\mu$. Different colors of the lines correspond to various numbers of WDM channel pairs, $N$, utilized by Alice and Bob, as shown in the legend. The transmission profiles of WDM channels is assumed to be either rectangular, given by Eq.\,(\ref{eq:rectchannel}) with the width of $\Delta\omega=2\pi\times50\,\mathrm{GHz}$ (solid lines), or Gaussian, given by Eq.\,(\ref{eq:Gausschannel}) with the standard deviation of $\sigma_f=2\pi\times50\,\mathrm{GHz}$ (dot-dashed lines). The central angular frequencies of the WDM channel pairs utilized by the trusted parties are defined according to Eq.\,(\ref{eq:omegasetnegative}). All the results are calculated for the transmittance of the channels connecting the source with the trusted parties equal to $T=10^{-3}$. (b) The ratio between the probability for only Alice's photon to enter the WDM channel belonging to the pair denoted by number $n$, $p_{+-}^\mathrm{Gaus}$, and the probability for both photons from an SPDC pair to enter this channel pair, $p_{++}^\mathrm{Gaus}$, given by the formulas (\ref{eq:pplusminusGaus}) and (\ref{eq:pplusplusGaus}) respectively, plotted as a function of the EPMF width for $\tau_{p}=333\,\mathrm{ps}$.}
\label{fig:KeyRateWDMprofile}
\end{figure}

The comparison between the gain function values calculated for the cases of rectangular and Gaussian profiles of the WDM channel transmission is presented in Fig.\,\ref{fig:KeyRateWDMprofile}\,(a). While the key generation rate that can be obtained just from the central WDM channel pair ($N=1$) is approximately independent on the EPMF width in the rectangular case, the situation is different when the WDM channel transmission profile is Gaussian. In this case the key generation rate for $\sigma_{cr}\rightarrow\infty$ reaches values that are approximately $50\%$ lower than for $\sigma_{cr}=100\,\mathrm{GHz}$. This decrease coincides with the increase of the ratio between the probability for only signal photon to enter the proper WDM channel, $p_{+-}^\mathrm{Gaus}$, and the probability for both signal and idler photons to enter the considered channel pair, $p_{++}^\mathrm{Gaus}$, as illustrated in Fig.\,\ref{fig:KeyRateWDMprofile}\,(b). It should be mentioned here that since the considered setup configuration is symmetric, the probability for only idler photon to enter a given WDM channel pair, $p_{-+}^\mathrm{Gaus}$ is always equal to $p_{+-}^\mathrm{Gaus}$. While the ratio $p_{+-}^\mathrm{Gaus}/p_{++}^\mathrm{Gaus}$ is negligible for $\sigma_{cr}=100\,\mathrm{GHz}$, it approximately reaches the value of $0.5$ for $\sigma_{cr}\rightarrow\infty$. However, the events in which only one of the SPDC photons enters a given WDM channel pair can contribute only to the random coincidences, in which case the probability for error is $50\%$. Thus, increasing $p_{+-}^\mathrm{Gaus}/p_{++}^\mathrm{Gaus}$ makes the QBER grow, lowering the obtainable key generation rate in comparison with the case of rectangular WDM channel profile, for which the ratio of $p_{+-}^\mathrm{rect}/p_{++}^\mathrm{rect}$ remains negligibly small for all the relevant pairs of values of $\sigma_{cr}$ and $\tau_p$.

It turns out that when the EPMF width becomes high enough for other WDM channel pairs to significantly contribute to the overall key generation rate, the ratio of $p_{+-}^\mathrm{Gaus}/p_{++}^\mathrm{Gaus}$  calculated for them is approximately the same as for the central channel pair. Therefore, $K_\mathrm{WDM}^\mathrm{total}$ that can be obtained by the trusted parties when using WDM modules with Gaussian transmission profile is about two times lower than in the case of rectangular profiles, regardless of how many channel pairs Alice and Bob use in total. For realistic WDM modules, with channel transmission profiles steeper than the Gaussian function, one should expect the gain function to take values in between the results for Gaussian and rectangular profiles presented in Fig.\,\ref{fig:KeyRateWDMprofile}\,(a).

\section{Discussion}
\label{Sec:Summary}

In this manuscript we considered the possibility for increasing the overall key generation rate produced by a DV QKD system utilizing SPDC source of entangled photon pairs by employing WDM modules. These devices can be used to split the signal and idler photons with different wavelengths into separate detection systems and perform multiple key generation processes in parallel. While employing WDM modules to enable simultaneous transmission of quantum and classical information in the existing telecommunication networks has been studied for many years, the idea to use them in order to enhance the performance of a single QKD system is quite recent. In this work we presented its first comprehensive theoretical analysis. In order to maximize the key rate provided by such setup configuration we performed optimization of the SPDC source over the EPMF width of the utilized nonlinear crystal and the intensity and duration of the pump laser pulses. The results were compared with the key rate produced by the traditional QKD system, showing the potential for significant improvement. Since low key generation rate is one of the main disadvantages of realistic DV QKD schemes in general, the outcomes of our analysis can have great importance for the practical quantum communication. They can help the experimenter to either choose the setup elements that would allow them to obtain secure key with desirable speed or to estimate the key generation rate that they can be able to get from a previously defined setup. They can also allow them to judge if it is worth using WDM modules in their scheme and how many WDM channels can possibly contribute efficiently to the overall key rate.

In the presented analysis we investigated both the cases of the SPDC source generating strongly negative and strongly positive spectral correlation between the signal and idler photons. For the negative correlation scenario the number of WDM channel pairs that can contribute to the overall key rate is determined primarily by the EPMF width, while the pump laser pulse duration is responsible for the key rate obtainable from individual WDM channel pairs. When the spectral correlation is positive, the correlated WDM channel pairs have to be defined in different way and the roles of the aforementioned parameters are reversed. We discussed the optimal values of these two quantities, considering currently available technology. The results for the overall key generation rate obtained in our analysis exhibit weak dependence over the WDM channel width. We investigated both Gaussian and rectangular transmission profile for these channels, showing approximately two times larger overall key generation rate for the latter case when the source parameters are optimized.

While in theory the potential increase of the overall key generation rate offered by the considered technique is limited only by the number of WDM channels available for Alice and Bob, several other issues may restrain it in realistic situations. First of all the value of the EPMF width for the nonlinear crystal used as the source of entangled photon pairs, that would be required for optimal performance of the QKD system with multiple WDM channel pairs, can be hard to achieve in practice. Although the topic is not well covered in the literature, the available EPMF width calculation results, regarding e.g. the very popular BBO crystal, show that this may certainly be an important issue. However, more detailed study of the phase matching function would be needed, especially for the case of the periodically poled type of crystals, which may offer bigger versatility in terms of $\sigma_{cr}$, before the importance of the aforementioned limitation can be fully assessed. In principle there could be an analogous issue related to the pump laser pulse duration. However, this parameter is much more versatile in practice and should be readily achievable as long as it does not have to be smaller than single femtoseconds. Meanwhile, the results of our calculation made for the positive spectral correlation case suggest that $\tau_p\approx10^{-13}\,\mathrm{s}$ would already be sufficient for the QKD system with at least ten WDM channel pairs to be fully used. 

Here it should be mentioned that the required optimal values of both the pump laser pulse duration and the EPMF width can be modified by changing the spectral separation between the individual WDM channels and their spectral width, as discussed in Sec.\,\ref{Sec:WDMproperties}.  This may help the trusted parties to adjust the properties of the WDM QKD scheme to their experimental capabilities and, in consequence, increase the potential gain offered by it. On the other hand, another practical limitation on the performance of the studied QKD setup configuration comes from the WDM channel profiles, which are never exactly rectangular, as assumed in most of our calculations. However, since their slopes are typically much steeper than the Gaussian function, we expect the reduction of the overall key rate caused by this issue to be rather minor. Finally, it is important to realize that WDM QKD setup studied in this paper requires either using separate detection systems for every WDM channel or utilizing detector time multiplexing (DTM) technique \cite{Fitch2003,Achilles2004,Kaltwasser2024} in order to be able to perform multiple key generation processes in parallel. In the former of these cases the entire QKD setup would become more expensive than the traditional setup configuration, with both Alice and Bob using single detection systems. However, it is worth pointing out that implementing WDM QKD would still require considerably less resources than generating multiple keys in parallel using totally independent QKD systems, for which not only multiple detectors, but also multiple sources and possibly separate quantum channels would have to be provided. On the other hand, utilizing DTM technique to reduce the number of required detectors could impose serious limitation on the repetition rate of the WDM QKD setup.

Although we focused our attention on the entanglement-based version of the BB84 protocol, the results of our work are relevant for broader class of schemes. The analysed method can have positive effect on the key rate as long as the states of Alice’s and Bob’s subsystems are spectrally entangled before the measurement leading to generation of the key bit takes place. If this is the case, by introducing WDM modules and establishing correlated WDM channel pairs, the large part of the uncorrelated photon pairs can be filtered out while the correlated pairs are sent to separate detection systems, enabling parallel generation of multiple secure keys. Several DV QKD protocols (\emph{e.g.}\,E91 \cite{Ekert1991}) or their entanglement-based counterparts (\emph{e.g.}\,SARG04 \cite{Scarani2004}, six-state protocol \cite{Bruss1998}), constructed by replacing a single-photon source with a suitable photon-pair source followed by an additional detection system, can satisfy this condition. Beside the polarization encoding, considered in our manuscript, it is also possible to apply different encoding variants, for example based on time bins \cite{Honjo2024}. However, in order for the WDM QKD method to work properly, the encoding should rather be done automatically during the process of photon pair generation, not later on by sending the already existing photons to some additional device (\emph{e.g.}\,phase modulator in the case of phase-encoding schemes). The reason for this requirement is that in the WDM QKD it is generally desirable to produce many pairs of photons in a single event. Applying random and independent encoding to all of them during their pass through the encoding device may turn out to be extremely difficult or even completely impossible to do in practice.

Unfortunately, the considered solution for increasing the key generation rate produced by DV QKD schemes does not appear to be suitable for improving the performance of the very popular class of measurement-device-independent (MDI) QKD protocols \cite{Lo2012,Lucamarini2018}. Obviously, one could think of adding WDM modules to the entanglement-based version of such scheme \cite{Xu2013} – one for each of the trusted parties and two for Charlie, in order to allow him to spectrally segregate photons received from Alice and Bob, respectively. Then, Charlie could define the WDM channels centered on the same wavelength as the correlated ones, and send the photons entering them to separate Bell measurement systems. Later on, during the postprocessing stage of the protocol, he would have to announce not only the measurement result, but also which of his systems performed the successful detection. If the spectral correlation between Alice’s photon and the photon she had send to Charlie was strong, this information would allow her to find out which of her own measurement systems detected the correlated photon and then check the result. An analogous action performed by Bob would finally allow the trusted parties to generate a secure bit. If many pairs of photons were produced by Alice’s and Bob’s SPDC sources in a single attempt, and Charlie announced successful detections in several of his measurement systems, Alice and Bob could in principle obtain many key bits at once. However, the discussed scenario has an important disadvantage. Since the photon send from Alice to Charlie and the photon send  from Bob to Charlie are uncorrelated with each other, the probability for them to enter any of the correlated WDM channel pairs would scale approximately as $1/N$, where $N$ is the number of the utilized channels per module. This would lower the key generation rate obtained from every pair of channels, cancelling the possible advantage provided by the fact that $N$ keys are generated in parallel. Additionally, it is worth pointing out that the twin-field variant of MDI QKD protocols \cite{Lucamarini2018} is essentially based on phase encoding, which would be unsuitable for the analysed WDM QKD method, as already indicated in the previous paragraph. 

The presented analysis was performed with several simplifying assumptions. In particular, only noiseless case was considered, with the ideal photon-number-resolving detectors used by the trusted parties. Therefore, the results are most relevant for the medium communication distance, such as can be found in typical metropolitan networks. In the future we plan to continue this research by investigating the influence of noise on the obtained results in various configurations of the detection systems and considering the effects of chromatic dispersion in typical single-mode fibers. Furthermore, in order to build the bridge between our work and Ref.\,\cite{Pseiner2021}, where the idea of utilizing WDM modules to increase the key generation rate of the QKD system was first proposed, we intend to compare the key generation rates that can be produced with the use of pulsed and continuous-wave pump lasers.

\section*{Acknowledgements}

The authors acknowledge financial support from the project ``Secure quantum communication in multiplexed optical networks'' run by the National Science Centre (NCN) in Poland as a part of the OPUS 20 + LAP programme (grant no.\,2020/39/I/ST2/02922). 
        
\section*{Appendix}

Basing on the theory presented in Ref.\,\cite{Mauerer2009} one can derive the following formula for the probability for a type-II SPDC source to generate $k$ photon pairs in a single polarization mode in total:
\begin{eqnarray}
    &\pi(k)&=\sum_{k_1=0}^{k}\sum_{k_2=0}^{k-k_1}\dots\sum_{k_{M-1}=0}^{k-k_1-\dots-k_{M-2}}\frac{\left(\mu\lambda_0\right)^{k_1}}{\left(\mu\lambda_0+1\right)^{k_1+1}}\frac{\left(\mu\lambda_1\right)^{k_2}}{\left(\mu\lambda_1+1\right)^{k_2+1}}\times\dots\times\frac{\left(\mu\lambda_{M-2}\right)^{k_{M-1}}}{\left(\mu\lambda_{M-2}+1\right)^{k_{M-1}+1}}\nonumber\\&&\times\frac{\left(\mu\lambda_{M-1}\right)^K}{\left(\mu\lambda_{M-1}+1\right)^{K+1}},
   \label{eq:nonpoissondistr}
\end{eqnarray}
where $\mu$ is the mean number of the produced pairs, $K=\sum_{s=1}^{M-1}k_s$ and $\lambda_m$ is the relative strength of the $m$-th mode, that can be expressed in terms of the parameters $\sigma_{cr}$ and $\tau_p$ as
\begin{equation}
   \lambda_m=\frac{8\sigma_{cr}\tau_p\left(2-\sigma_{cr}\tau_p\right)^{2m}}{\left(2+\sigma_{cr}\tau_p\right)^{2m+2}}.
   \label{eq:lambda}
\end{equation}
In principle the summation in Eq.\,(\ref{eq:nonpoissondistr}) should be performed over infinite number of modes. However, because $0<\lambda_{m+1}/\lambda_{m}<1$, the contribution of the modes with numbers higher than $M$ to the probabilities $\pi(k)$ becomes negligible provided that sufficiently large value of $M$ is taken. Unfortunately, since the EPMF widths and pump laser pulse durations relevant for our investigation are such that $\sigma_{cr}\tau_p\gg1$ ($\sigma_{cr}\tau_p\ll1$) for the case of strongly negative (positive) spectral correlation considered in Sec.\,\ref{Sec:WDMnegative} (Sec.\,\ref{Sec:WDMpositive}), the number of modes contributing to the real photon pair statistics is large. Thus, the numerical calculations of $K_\mathrm{WDM}^\mathrm{total}$, needed to generate the results presented in the main body of this manuscript, could not be done effectively with the use of Eq.\,(\ref{eq:nonpoissondistr}). However, in the limit of infinitely strong spectral correlation, \emph{i.e.} when either  $\sigma_{cr}\tau_p\rightarrow\infty$ or $\sigma_{cr}\tau_p\rightarrow0$, the statistics of the produced photon pairs take the form of Poisson probability distribution \cite{Mauerer2009}. Therefore, we performed all the aforementioned calculations with the probabilities $\pi(k)$ approximated by Poisson statistics. 

\begin{figure}[tb]
\centering
\includegraphics[width=0.8\linewidth]{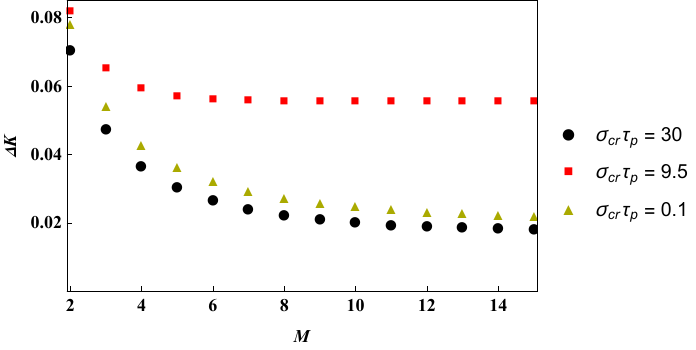}
\caption{Relative detuning of the key generation rate caused by approximating the real statistics of photon pairs emitted by the SPDC source in a single polarization mode by the Poisson probability distribution, defined in Eq.\,(\ref{eq:DeltaK}), plotted as a function of the number of modes $M$ considered in the calculation of the real statistics. Different colors and styles of the points correspond to different values of the $\sigma_{cr}\tau_p$ product, as indicated in the legend. The plots are made for only the single WDM channel pair utilized by Alice and Bob ($N=1$).}
\label{fig:PoissonApprox}
\end{figure}

In order to assess the impact of this approximation on the presented results let us define the relative detuning of the key generation rate
\begin{equation}
    \Delta K(M)=\frac{K_\mathrm{WDM,M}^\mathrm{total}-K_\mathrm{WDM,Poisson}^\mathrm{total}}{K_\mathrm{WDM,Poisson}^\mathrm{total}},
    \label{eq:DeltaK}
\end{equation}
where $K_\mathrm{WDM,M}^\mathrm{total}$ is the overall key rate calculated when utilizing the photon pair probability distribution given by Eq.\,(\ref{eq:nonpoissondistr}) with $M$ modes involved, while $K_\mathrm{WDM,Poisson}$ is the analogous result obtained with the use of Poisson probability distribution. The relative detuning of the key generation rate is an asymptotically decreasing function of $M$, as can be seen in Fig.\,\ref{fig:PoissonApprox}, where it is plotted for the case of $N=1$ and three different values of the $\sigma_{cr}\tau_p$ product. The results do not depend on the number of WDM channel pairs considered in this calculation. For $\sigma_{cr}\tau_p\approx9.5$, which is the smallest value relevant for our analysis in the case of negative spectral correlation, $ \Delta K(M)\lim_{M\to\infty}0.055$. The value of this function drops below $0.02$ for $\sigma_{cr}\tau_p=30$. Corrections of $K_\mathrm{WDM}^\mathrm{total}$ of this order would be barely visible in the figures regarding negative spectral correlation case presented in the main body of this manuscript. Furthermore, the difference would become even smaller for the pairs of values of the EPMF width and the pump laser pulse duration allowing for substantial overall key rate improvement with use of the considered WDM method. Similar verdict can be delivered in the case of positive spectral correlation, where the largest value of $\sigma_{cr}\tau_p$ relevant to our calculation is $0.1$. In this situation $\Delta K(M)\lim_{M\to\infty}0.021$, as can be also seen in Fig.\,\ref{fig:PoissonApprox}. Therefore, the assumed Poisson approximation of the statistics of photon pairs emitted by the SPDC source does not influence any of the conclusions drawn in this manuscript.

\bibliography{MultiplexQKDRefList}

\end{document}